\documentclass{webofc}
\usepackage[varg]{txfonts}   
\def\bra#1{\left\langle #1\right|}
\def\ket#1{\left| #1\right\rangle}

\def\PRL{Phys.\,Rev.\,Lett.}
\woctitle{QCD@Work 2014 
}
\begin{document}
\title{On the {\it $\Delta A_{\rm CP}$ saga}\thanks{
The phrase {\it $\Delta A_{\rm CP}$ saga} was used for the first time by Guy Wilkinson at 
Beauty 2013 \cite{saga1}, see also \cite{saga2}.}
}

\author{Pietro Santorelli\inst{1,2}\fnsep\thanks{\email{pietro.santorelli@na.infn.it}} 
}
\institute{Dipartimento di Fisica, Universit\`a di Napoli
Federico II, Complesso Universitario di Monte S. Angelo,
Via Cintia, Edificio 6, 80126 Napoli, Italy
\and
Istituto Nazionale di Fisica Nucleare, Sezione di Napoli, Napoli, Italy
}

\abstract{
We discuss a model in which the large SU(3) flavor violations in singly Cabibbo 
suppressed decays of neutral $D$ mesons are ascribed exclusively to the final state interactions. 
The agreement with the experimental data on the branching ratios is obtained with 
large strong phase differences which are also necessary for substantial direct CP violation. 
While the value of the CP violating asymmetries depends on the strength 
of the penguin contribution, we predict an asymmetry for the decays into charged 
pions more than twice larger than that for charged kaons and having opposite sign. 
}
\maketitle
\section{Introduction}
\label{intro}
At the end of 2011 LHCb Collaboration measured  the difference in the CP-violating asymmetries 
between the two decay channels  $D \to K^+ K^-$ and $D \to \pi^+\pi^-$ obtaining \cite{LHCb1}
\begin{equation}
\Delta A_{\rm CP}= A_{\rm CP}(K^+ K^-) - A_{\rm CP}(\pi^+ \pi^-) =
(-0.82 \pm  0.21 \pm 0.11)\%\,.
\end{equation}
This result was confirmed by the measurements of CDF \cite{CDF} 
and Belle \cite{Belle} Collaborations
\begin{equation}
\Delta A_{\rm CP}  =  (-0.62 \pm  0.21 \pm 0.10)\%\,,\nonumber\\
\end{equation}
\begin{equation}
\Delta A_{\rm CP} =  (-0.87 \pm  0.41 \pm 0.06)\%\,,
\end{equation}
respectively.
These experimental results on direct CP violation in the Singly Cabibbo Suppressed (SCS)
charm decays produced a large amount of interest. In the Standard Model, indeed,  
CP Violation (CPV) in charm decays is commonly expected  to be very small 
\cite{GrossmanKaganNir} and so CP asymmetries of the order of one percent could be a 
signal of New Physics. The theoretical community have interpreted these results 
by considering them as a sign of New Physics \cite{NewPhys} or compatible with the 
Standard Model \cite{StanMod,Silvestrini}. We prefer the latter hypothesis \cite{our}.
Meanwhile,  new results by LHCb, using more data and different way to identify the 
flavour of the charmed meson, show that, at the level of $10^{-3}$,  there is no 
evidence of CP violation in  $D^0 \to K^+K^-,\pi^+\pi^-$  decays 
\cite{LHCbLast}
\begin{equation}
\Delta A_{\rm CP} = (+0.14 \pm 0.16 \pm 0.08 )\% .
\end{equation}
Nevertheless, it is interesting to give an answer to the question regarding the amount 
of the direct CP violation in non-leptonic SCS D decays in the 
Standard Model. In the following we will show the main ideas and the results obtained 
in our paper \cite{our}.

\section{The model}
Many years ago we  presented a calculation of the decay  branching ratios of $D$ and $D_{\rm s}$ 
mesons \cite{ourold} based on factorization hypothesis and a model to account for rescattering effects through nearby resonances. The results were in reasonably good agreement with the allowed experimental data and predicted CP asymmetries at least one order of magnitude smaller than what was found in \cite{CDF,LHCb1}. The experimental data however did change in the meanwhile and 
therefore we have carried out a new analysis by considering only the SCS decays \cite{our}. 
In \cite{ourold} we stressed the fact that 
the observed SU(3) flavor violations were essentially due to the rescattering effects. 
In \cite{our}, we evaluate the weak decay amplitudes assuming SU(3) symmetry 
so that we can consider the rescattering later. Furthermore, we approximate the hamiltonian for $D$ weak decays with its $\Delta U = 1$ part when estimating branching ratios, introducing the $\Delta U = 0$ terms only for the calculation of asymmetries. This is a reasonable approximation because of the smallness of the relevant CKM elements, 
$\left |V_{\rm ub} V_{\rm cb}^*\right | \ll |V_{\rm ud(s)} V_{\rm cd(s)}^\ast|$.

\subsection{Decay branching ratios}
The weak effective hamiltonian for SCS charmed particles is
\begin{equation}
\label{Hamilt}
\mathcal{H}_{\rm w} = \frac{G_F}{\sqrt 2} V_{ud}\,V_{cd}^*
\;[C_1 Q_1^d+C_2 Q_2^d] \;
+ {{G_F}\over{\sqrt 2}} V_{us}\,V_{cs}^*\;[C_1 Q_1^s+C_2 Q_2^s] \;
- {{G_F}\over{\sqrt 2}} V_{ub}\,V_{cb}^*\;\sum_{i=3}^6\;C_i Q_i + h.c. \,,
\end{equation}
where the expressions for the operators $Q_i^j$ and the numerical values used for the Wilson 
coefficients $C_i$ can be found in \cite{our}. Considering the transformation properties 
respect to the U-spin, the effective hamiltonian can be decomposed in two parts
\begin{eqnarray}
H_{\Delta U = 1} &=& {{G_F}\over{2\,\sqrt 2}} 
\underbrace{(V_{us}\,V_{cs}^*-V_{ud}\,V_{cd}^*)}_{\sin \theta_C \cos \theta_C }
\left [C_1( Q_1^s-Q_1^d)+C_2 (Q_2^s-Q_2^d)\right ]\,, \\
\label{Uzero}
H_{\Delta U = 0}  & = & - \; {{G_F}\over{\sqrt 2}} 
V_{ub}\,V_{cb}^*\;\left\{\sum_{i=3}^6\;C_i Q_i + 
{1 \over 2}\left [C_1( Q_1^s+Q_1^d)+C_2 (Q_2^s+Q_2^d)\right ]\right\}\;.
\end{eqnarray}
Neglecting $H_{\Delta U = 0}$ and observing that the neutral charmed meson 
$D^0$ is U-spin singlet it can be easily demonstrate that the SCS decays can be written in 
terms of only two independent amplitudes. In fact, there are two independent 
combinations of $S$-wave states having $U$=1
\begin{eqnarray}
|v_1> &=&{1\over 2}\Big\{|K^+\,K^-> +|K^-\,K^+>-|\pi^+\,\pi^->-|\pi^-\,\pi^+>\Big\}\;, 
\nonumber\\
|v_2> &=& {\sqrt{3} \over {2 \sqrt{2}}} \Big\{|\pi^0 \, \pi^0>-|\eta_8\,\eta_8>-{1 \over \sqrt{3}}(|\pi^0\,\eta_8>+
 |\eta_8\,\pi^0>)\Big\}\; , 
 \end{eqnarray}
that may be rewritten in terms of the following two states with given transformation properties  
under SU(3)
\begin{eqnarray}
\label{8SU3}
|8,U=1> & = & \frac{\sqrt{3}} {2 \sqrt{5}} 
\left\{\frac{}{}\! |K^+K^-> +|K^-K^+>-|\pi^+\pi^->-|\pi^-\pi^+> \right. \nonumber\\
 &&\hspace{1cm} 
 - \left. \left[|\pi^0\pi^0>-|\eta_8\eta_8>-{1 \over \sqrt{3}}(|\pi^0\eta_8>+
 |\eta_8\pi^0>)\right]\right\},  \\
 \label{27SU3}
 |27,U=1> & = & \frac {1}{\sqrt{10}} 
 \left\{\frac{}{}\! |K^+K^-> +|K^-K^+>-|\pi^+\pi^->-|\pi^-\pi^+>
 \right.  \nonumber\\
  && \hspace{1cm}
  + \left. {3 \over 2}\,\left[|\pi^0\pi^0>-|\eta_8\eta_8>-{1 \over \sqrt{3}}(|\pi^0\eta_8>+
 |\eta_8\pi^0>)\right]\right\}.
 \end{eqnarray}
Using these states, the two independent amplitudes can be written in terms of the diagrammatic 
ones, color connected ($T$) and color suppressed ($C$), in the following way
\begin{equation}
\bra{8, U=1}H_{\Delta U = 1}\ket{D^0} \propto T-\frac{2}{3}C\,,
\hspace{1cm}
\bra{27, U=1}H_{\Delta U = 1}\ket{D^0} \propto T+C\,. \nonumber
\end{equation}
To test the model with the experimental data on the branching ratios we should  consider the final state interactions. The experimental data strongly violate SU(3) symmetry. In fact, in the limit of SU(3) flavor symmetry the relation  $A(D^0\to K^+\, K^-) = -A(D^0\to \pi^+\, \pi^-)$ holds, but the experimental data on the branching ratios violate the 
previous relation \cite{PDG}
\begin{eqnarray}
Br(D^0\to \pi^+\pi^-) & = &  (1.402\pm 0.026)\times 10^{-3}\;,\nonumber\\ 
Br(D^0\to K^+ K^-)    & = & (3.96 \pm0.08)\times 10^{-3}\;. \nonumber 
\end{eqnarray}

In our model the SU(3) breaking is given by the final state interactions, described as the effect of 
resonances in the scattering of the final particles. 
The possible resonances have the following SU(3) and isospin quantum numbers:
$(8, I=1)$, $(8,I=0)$ and $(1,I=0)$. The two states with $I=0$ can mix, providing two resonances
\begin{eqnarray}
|f_0> &=& \sin \phi  \;\; |8,I=0> + \cos \phi \;\; |1,I=0> , \\
|f'_0> &=& - \cos \phi  \;\; |8,I=0> + \sin  \phi \;\; |1,I=0>. 
\end{eqnarray}
The angle $\phi$, the strong phases $\delta_0$, $\delta'_0$ and $\delta_1$ (corresponding to $f_0$, $f'_0$ and to the particle with isospin one, respectively)
and the two weak decay amplitudes are the free parameters of our model. We choose to fix  these parameters by comparing model predictions with experimental data on the 
SCS branching ratios.  In table~\ref{tab} we give the values of the free parameters: for a detailed discussion see \cite{our}. 

\begin{table}
\centering
\caption{The fitted value for the free parameters.}
\label{tab}       
\begin{tabular}{ccccc}
\hline
$C/T$ & $\phi$ & $\delta_0$ & $\delta^\prime_0$ & $\delta_1$ \\\hline
$-0.529$ & $+0.389$ & $\pm 2.58$  & $\pm 0.917$ &  $\pm 1.44$  \\\hline
\end{tabular}
\end{table}

\subsection{CP asymmetries}
The direct CPV occurs when the decay amplitudes for CP conjugate processes into final states $f$ and 
$\bar f$ (in our case $f\equiv \bar f$) are different in modulus. This requires the presence of at least 
two interfering decay amplitudes with different weak and strong phases. In our approach, the second amplitude is provided by the matrix elements of the $\Delta U = 0$ hamiltonian, eq.~(\ref{Uzero}), that contains both $Q_{1(2)}$ and "penguin" operators. For $H_{\Delta U = 0}$, the independent states 
are the following
\begin{eqnarray}
\label{Unull}
&&{1\over 2}\left\{|K^+\,K^-> +|K^-\,K^+>+|\pi^+\,\pi^->+|\pi^-\,\pi^+>\right\}\; ,   \nonumber\\
&&{1 \over 4} \left\{3\,|\pi^0 \, \pi^0>+|\eta_8\,\eta_8>+\sqrt{3}\,\left(|\pi^0\,\eta_8>+
 |\eta_8\,\pi^0>\right)\right\}\;  , \nonumber \\
 &&{1 \over \sqrt{3}} \left\{\frac{1}{4}|\pi^0 \, \pi^0>+\frac{3}{4}\,|\eta_8\,\eta_8>-\frac{\sqrt{3}}{4}\,(|\pi^0\,\eta_8>+
 |\eta_8\,\pi^0>)+  |K^0\,\bar{K^0}>+|\bar{K^0}\,K^0>\right\}\; . 
 \end{eqnarray}
As an example, we report the expressions for the second amplitude 
(the amplitude B \cite{our}, see also eq.~(\ref{amp})) for the  
decay modes of interest
\begin{eqnarray}
\label {pipiUz}
B(D^0 \to \pi^+ \pi^-) & = &  \left(P+\frac{T'}{2}\right)
\left\{\frac{1}{2} \left(e^{\imath \delta_0} +   e^{\imath \delta^\prime_0}\right)
  +  \left( - \frac{1}{6} \cos(2 \phi) - \frac{7}{4\sqrt{10}} \sin(2 \phi)\right) 
 \left(e^{\imath \delta^\prime_0} - e^{\imath \delta_0}\right) \right\} \nonumber \\
&+& \left( T' + C'  \right) \; \left\{ \frac{3}{20} -
 \frac{3}{40}\; \left(e^{\imath \delta_0} +   e^{\imath \delta^\prime_0}\right)  \right. \nonumber\\
&+& \left. \left[ {1 \over 120} \, \cos (2 \phi) + {1\over {4 \sqrt {10}}}\,\sin (2 \phi) \right] \;
 \left(e^{\imath \delta^\prime_0} - e^{\imath \delta_0}\right) \right\} \;, \\
\label{KKUz}
 B(D^0\to K^+K^-) & = &  \left(P+\frac{T'}{2}\right)
 \left\{
 \frac{1}{4} \left(e^{\imath \delta_0} +   e^{\imath \delta^\prime_0}\right)  
  +   \left( -\frac{5}{12} \cos(2 \phi) + \frac{1}{4\sqrt{10}} \sin(2 \phi)\right) 
\left(e^{\imath \delta^\prime_0} - e^{\imath \delta_0}\right)\right.\nonumber\\
 && \left.+  \frac{1}{2} e^{\imath \delta_1}\right\}\nonumber \\
 & + & 
 \left( T'+C' \right) \left\{\frac{3}{20}\; - \frac{1}{40}\; \left(e^{\imath \delta_0} +   
 e^{\imath \delta^\prime_0}\right) + \frac{7}{120} \, \cos (2 \phi)
 \left(e^{\imath \delta^\prime_0} - e^{\imath \delta_0}\right) \right.\nonumber\\
 &&  \left. - \frac{1}{10}\, e^{\imath \delta_1} \right\}.
 \end{eqnarray}
Writing the general expressions for the decay amplitudes $D\to f$ and its CP-conjugate
\begin{eqnarray}
{\mathcal A}(f) & = & A \; e^{\imath \delta_A} + B \; e^{\imath \delta_B}\:,\nonumber\\
\label{amp}
{ \bar {\mathcal A}}(\bar f)  & = &A^* \; e^{\imath \delta_A} + B^* \; e^{\imath \delta_B}\: ,
\end{eqnarray}
respectively, the CP asymmetry is given by
\begin{equation}
\label{asym}
A_{\rm CP}(f) = {{|{\mathcal A}(f)|^2-|{ \bar {\mathcal A}(\bar f)}|^2} \over {|{\mathcal A}(f)|^2+
|{ \bar {\mathcal A}(\bar f)}|^2}} =  {{2\; \Im (A^*\,B)\; \sin (\delta_A - \delta_B)} \over
{|A|^2 + |B|^2 + 2 \; \Re (A^*\,B)\; \cos (\delta_A - \delta_B)}}\:,
\end{equation}
where $\delta_{A(B)}$ are the strong phases and the amplitudes A and B contain the weak phases.
In eqs.~(\ref{pipiUz},\ref{KKUz}) $P$ represents the contribution of the penguin diagram, while the 
terms $T'$ and $C'$ are related to $T$ and $C$ by the following relations
 \begin{equation}
 \label{Dparam}
 T'= - \; T \; \frac{V_{ub}\, V_{cb}^*}{\sin\theta_C \cos \theta_C}\quad {\rm and} \quad
 C'= - \; C \; \frac{V_{ub}\, V_{cb}^*}{\sin\theta_C \cos \theta_C}\;.
 \end{equation}
Neglecting the $T'$ and $C'$ contributions,  the amplitudes and thus the asymmetries get a simplified expression. If we consider, for example, the $K^+K^-$ final state we have
\begin{equation}
{\mathcal A}(K^+K^-) \simeq T\; f_T(\delta_i,\phi,C/T)\;+\; P\;f_P(\delta_i,\phi) \;,
\end{equation}
and the CP-asymmetry is 
\begin{equation}
A_{\rm CP}(K^+K^-) \simeq \frac {2 \; \Im(f_T\, f_P^*)}{|f_T|^2}\;\;  \frac {\Im (P)}{T}\;,
\end{equation}
where 
\begin{equation} 
\label{penguin}
 \frac{\Im (P)}{T} = \frac{|V_{ub}\,V_{cb}|}{\sin \theta_C \cos \theta_C} \sin \gamma 
 \frac{<K^+\,K^-| \; \sum_{i=3}^6\;C_i Q_i + {1 \over 2}[C_1\{Q_1^s+Q_1^d\}+C_2 \{Q_2^s+Q_2^d\}]
  \;| D^0>}
 {<K^+\,K^-|\,C_1( Q_1^s-Q_1^d)+C_2 (Q_2^s-Q_2^d)\,|D^0>} 
 = 6.3\,10^{-4} \kappa \; .
\end{equation}
In eq.~(\ref{penguin}), $\{Q_i\}$ indicates the penguin contraction of the operator $Q_i$.  If we choose 
the lowest values in each column of table~\ref{tab} the asymmetries are 
\begin{equation}
A_{\rm CP}(K^+ K^-) =  + 1.469 \; \frac{\Im (P)}{T}\; , 
\end{equation}
\begin{equation}
A_{\rm CP}(\pi^+ \pi^-) = - 3.362 \; \frac{\Im (P)}{T}.
\end{equation}
Notice that our choice is related to the fact that the resonance  $f_0$(1710), the one with
lower mass, prefers to decay into kaons \cite{PDG} and thus it should be recognized  as  $f'_0$.\\
Putting all together we have
\begin{equation} 
\Delta A_{\rm CP} =  3.03 \; 10^{-3} \kappa \; .
 \end{equation}
Thus Aasymmetries of the order of percent can be obtained with a value of $\kappa$ around three. 
As far as the sign of  $\Delta A_{\rm CP}$ is concerned, $\kappa$ and  $\Delta A_{\rm CP}$ are 
negative in the factorization approximation, in agreement with the majority of experimental results.
We note that if factorization is used a considerably smaller value for $\kappa$ would be expected, due to the littleness of the Wilson coefficients of QCD penguin operators. However, large penguin contribution 
and penguin contraction of the current-current operators could give a value of $\kappa$ compatible with
CP asymmetries of the order of percent (as in the paper of J. Brod, A.L. Kagan and J. Zupan in \cite{StanMod}).  We recall that the penguin diagrams were introduced many years ago in \cite{SVZ}
as a possible explanation of the ``octet enhancement''. A large contribution of these operators could successfully describe both the hyperon and the kaon non--leptonic decays. For a comprehensive
discussion of the status of  ``$\Delta I = 1/2$ rule'' see, for example, \cite{Buras}.

\section{Conclusions}
We studied the singly Cabibbo suppressed decays of the neutral $D$ mesons in a model that ascribes 
the large SU(3) violations to final state interactions. We were able to reproduce the experimental data 
on the branching ratios and we have shown that CP violation asymmetries of the order of percent
are compatible with the Standard Model if we assume an enhancement of the penguin diagrams as 
in the case of the non-leptonic decays of kaons.

\begin{acknowledgement}
It is a pleasure to thank Franco Buccella, Maurizio Lusignoli and Alessandra Pugliese 
for a very enjoyable collaboration on the charm physics. I would like to thank Alexis Pompili and Vincenzo Vagnoni for discussions. Finally, I would like to thank the organizers of this workshop for the 
warm hospitality.
\end{acknowledgement}

\end{document}